\documentclass[final]{aipproc}

\layoutstyle{6x9}

\begin{document}

\title{Multiparton Tomography of Hot and Cold Nuclear Matter}

\author{Ivan Vitev}{
  address={Department of Physics and Astronomy, Iowa State University, 
Ames,~IA~50010}
}

\begin{abstract}
Multiple parton interactions in relativistic heavy ion reactions 
result in transverse momentum diffusion and medium 
induced non-Abelian energy loss of the hard probes traversing cold 
and hot nuclear matter. A systematic study of the interplay 
of nuclear effects on the $p_T \geq 2$~GeV inclusive hadron spectra 
demonstrates that the competition between nuclear shadowing, 
multiple scattering  and jet quenching leads to distinctly different 
enhancement/suppression of moderate and high-$p_T$ hadron production 
in $d+Au$ and $Au+Au$ collisions at RHIC. The associated increase 
of di-jet acoplanarity, measured via the broadening of the back-to-back
di-hadron correlation function, provides an additional experimental 
tool to test the difference in the dynamical properties of the media 
created in such reactions.
\end{abstract}

\maketitle

\subsection{Nuclear modification of hadron production}

Particle production from  a single hard scattering with 
momentum exchange much larger than 1/fm is localized in space-time. 
It is multiple parton scattering before 
or after the hard collision that is sensitive to the properties of the 
nuclear matter~\cite{Vitev:2002pf,Arleo:2002kh}. By comparing 
the high-$p_T$ observables
in $p+p$, $p+A$ and $A+A$ reactions, we are able to study the strong 
interaction dynamics of QCD in the vacuum, cold nuclear matter and hot
dense medium of quarks and gluons, respectively.  
So far the first two integral moments $\int dz \; z^n \rho(z)$  
of the matter density in the interaction region can be deduced from 
experimental measurements since they are related to the 
broadening~\cite{Qiu:2003pm,Luo:ui} 
and energy loss~\cite{Gyulassy:2000fs,Gyulassy:2003mc}
of a fast parton traversing nuclear matter.
 From~\cite{Qiu:2003pm,Gyulassy:2000fs} around midrapidity:
\begin{eqnarray}
\langle \Delta {\bf k}_T^2 \rangle  \!\!\!\!  &\approx& 
 \!\!\!\!  2\xi \int dz \, \frac{\mu^2}{\lambda_{q,g}} 
= 2 \xi \int dz \,  \frac{3  C_R \pi \alpha_s^2}{2} \rho^g(z) 
= \left\{ \begin{array}{ll}   2 \xi \,  
\frac{3  C_R \pi \alpha_s^2}{2} \, \rho^g \langle L \rangle \, , 
&  {\rm static} \\[1ex]
 2\xi \, \frac{3  C_R \pi \alpha_s^2}{2} 
\frac{1}{A_\perp} \frac{dN^{g}}{dy}
\,   \ln \frac{  \langle L  \rangle }{\tau_0}\, , & 1+1D  
\end{array}  \right. \nonumber \\
&&
\label{broad}
\end{eqnarray}
\vspace*{-1.cm}
\begin{eqnarray}
\langle \Delta E \rangle \approx  \int dz \, 
\frac{C_R\alpha_s}{2} \frac{\mu^2}{\lambda_{g}} \, z \, 
\ln \frac{2E}{\mu^2 \langle L \rangle }   
&=& \int dz \,  \frac{9 C_R \pi \alpha_s^3}{4} \rho^g(z) 
\, \ln \frac{2E}{\mu^2 \langle L \rangle }  \nonumber  \\
&& \qquad \quad  \!\! = \left\{ \begin{array}{ll}  
\frac{9 C_R \pi \alpha_s^3}{8} \, \rho^g \langle L \rangle^2 
\, \ln \frac{2E}{\mu^2 \langle L \rangle } 
 \, ,  &  {\rm static} \\[1ex]
 \frac{9  C_R \pi \alpha_s^3}{4} 
\frac{1}{A_\perp}  \frac{dN^{g}}{dy} \langle L  \rangle 
\,   \ln \frac{2E}{\mu^2 \langle L \rangle}  
 \, , & 1+1D  
\end{array}  \right.  \nonumber \\
&&
\label{deltae}
\end{eqnarray}
In Eq.(\ref{broad}) the factor 2 comes from 2D diffusion,  
$\xi \simeq {\cal O}(1)$ and $\rho^g$ is the  effective gluon density. 
For the 1+1D Bjorken expansion scenario $A_\perp$ is the 
transverse area of the interaction region, $\tau_0$ is the initial 
equilibration time and  $dN^g/dy$ is the effective gluon rapidity
density. In Eq.(\ref{deltae}) the dominant logarithmically 
enhanced contribution to mean energy loss computed in the GLV 
approach~\cite{Gyulassy:2000fs} is shown. For further discussion 
on medium induced radiative energy loss in QCD 
see~\cite{Gyulassy:2000fs,Gyulassy:2003mc}.   

\begin{figure}
  \includegraphics[width=2.7in,height=2.8in]{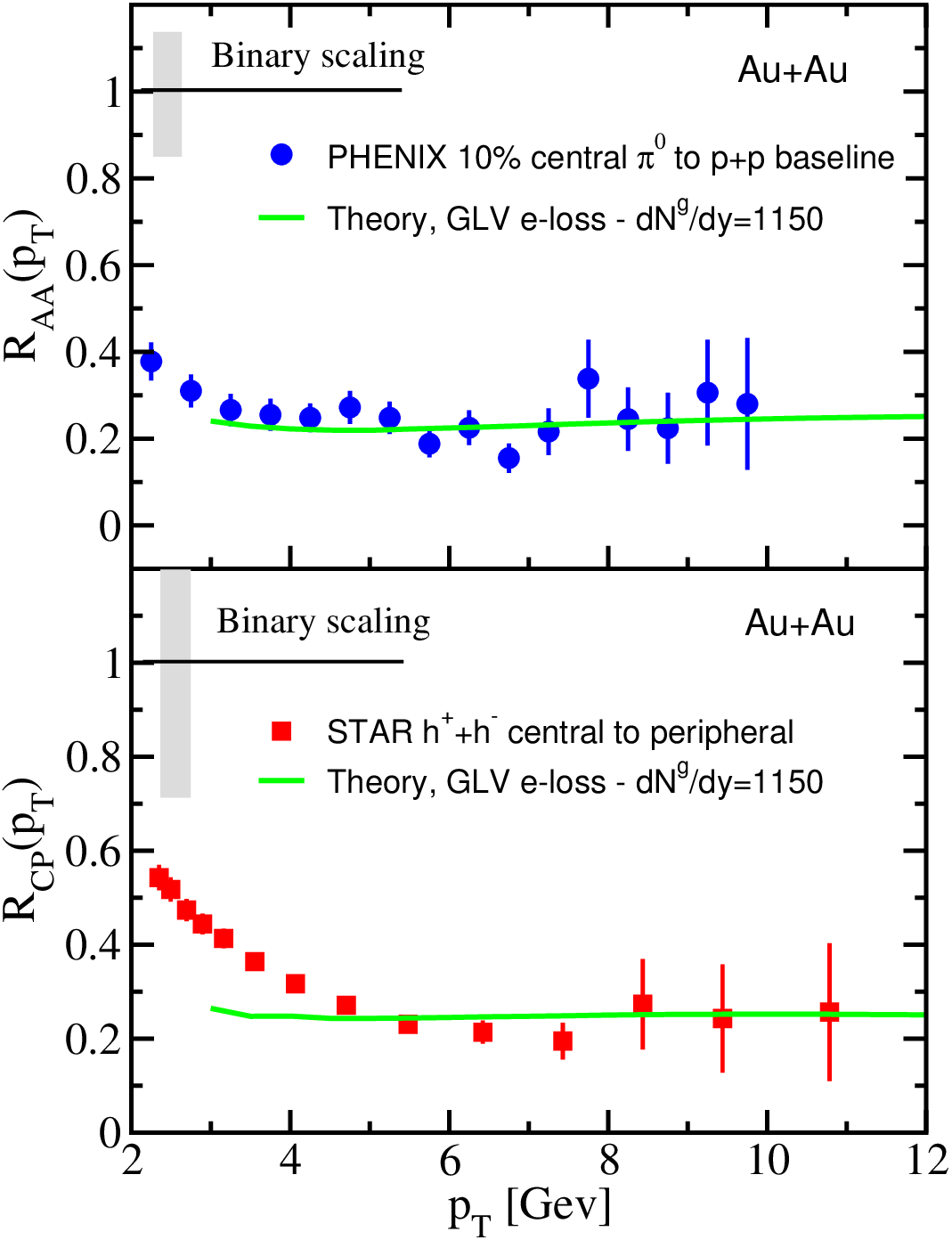}
\hspace*{0.3cm}
  \includegraphics[width=2.7in,height=2.8in]{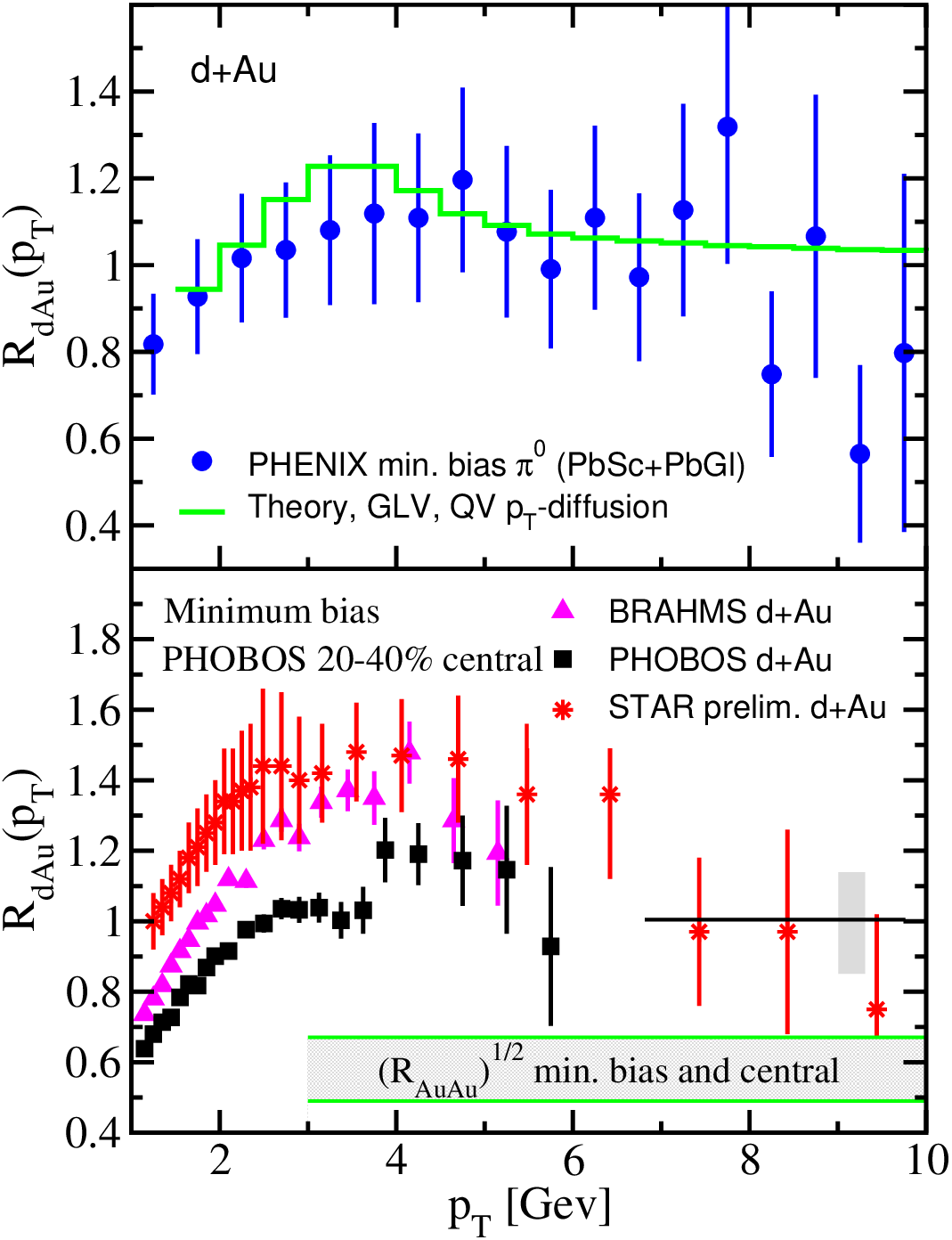}
\vspace*{0.3cm}
  \caption{Left panel: predicted suppression for $\pi^0$ 
and $h^+ + h^-$~\cite{Vitev:2002pf} in $Au+Au$ compared 
to PHENIX and STAR data~\cite{Adler:2003qi}. Similar quenching
is found by BRAHMS and PHOBOS~\cite{Adler:2003qi}.
 Right panel: predicted  small enhancement of neutral pions 
in $d+Au$~\cite{Adler:2003ii}, data is from PHENIX. 
Left bottom panel: a test of a suggested 
interpretation of high $p_T$-hadron suppression as result of 
initial state wavefunction modification ($R_{dAu} \approx
\sqrt{R_{AuAu}}$). Data is from
BRAHMS, PHOBOS and STAR~\cite{Adler:2003ii}.}
\label{fig1}
\end{figure}

Initial state parton broadening, nuclear shadowing and jet energy 
loss  are incorporated in the lowest order pQCD hadron production 
formalism as in~\cite{Vitev:2002pf}. The interplay of 
dynamical nuclear effects can be studied through the nuclear 
modification ratio
\begin{equation}         
R_{AB}(p_T)  = 
\frac{d N^{AB}}{dyd^2{\bf p}_T}  / 
\frac{T_{AB}(b) \; d \sigma^{pp}} {dyd^2{\bf p}_T}\; , \qquad
{\rm  about \; impact \; parameter}\; b \; {\rm in \;}A+B \;. 
\label{rba}
\end{equation} 
Figure~\ref{fig1} compares the predicted~\cite{Vitev:2002pf} 
approximately constant suppression of $\pi^0$ and $h^+ + h^-$ 
in $\sqrt{s} = 200$~AGeV  $Au+Au$ collisions at RHIC to 
PHENIX and STAR data~\cite{Adler:2003qi}. 
The overall quenching magnitude and its centrality dependence are set 
by $(\langle L \rangle / A_\perp) dN^g/dy \propto N_{part}^{2/3}$, 
$dN^g/dy=1150$.  The shape of $R_{AuAu}$ is a result of 
the interplay of all three nuclear effects. The full numerical 
calculation takes into account the dynamical Bjorken 
expansion of the medium, finite kinematic bounds, higher
order opacity corrections and approximates multiple gluon 
emission by a Poisson distribution~\cite{Vitev:2002pf,Gyulassy:2000fs}. 
In the right panel the Cronin enhancement, resulting from initial 
state parton broadening, $\langle \Delta {\bf k}_T^2 \rangle -
\langle \Delta {\bf k}_T^2 \rangle_{vac} \sim  
({\mu^2}/{\lambda}_g)_{eff} 
\langle L \rangle $~\cite{Vitev:2002pf,Qiu:2003pm} is seen to 
compare qualitatively to the shape of the 
PHENIX $\pi^0$ measurement~\cite{Adler:2003ii} in $d+Au$.
The calculations in Fig.~\ref{fig1} use the value 
$({\mu^2}/{\lambda}_g)_{eff} 
= 2 \times 0.14$~GeV$^2$/fm for the cold nuclear matter 
transport coefficient  constrained from existing Cronin 
data~\cite{Vitev:2002pf}, although somewhat smaller 
scattering strength may be favored by PHENIX data.  
Larger enhancement of $h^++h^-$ production, consistent 
with results form low energy $p+A$ measurements, is also 
shown~\cite{Adler:2003ii}.  The lower right 
panel rules out the scenario for the initial wavefunction  
origin of moderate and high-$p_T$ hadron suppression (see 
left panel of Fig.~\ref{fig1}) since in this case 
$R_{dAu} \approx \sqrt{R_{AuAu}}$. For further discussion on
the Cronin effect see~\cite{Vitev:2002pf,Zhang:2003jr}.

\subsection{Broadening of the away-side di-hadron correlation function}

The total vacuum+nuclear induced broadening for the two partons
in a  plane perpendicular to the collision axis in $p+A$ ($A+A$) 
reads~\cite{Qiu:2003pm}:
\begin{equation}
 \langle  {\bf k}_T^2 \rangle  = 
\langle  {\bf k}_T^2 \rangle_{vac} +   
\begin{array}{c} 1_{jet} \\ 
(2_{jets})
\end{array} 
\, \left( \frac{\mu^2}{\lambda} \right)_{eff} 
\langle L \rangle_{IS}  
+  2_{jets}\,  \left(\frac{1}{2} \right)_{projection}  \,
\left( \frac{\mu^2 }{\lambda} \right)_{eff} 
\langle L \rangle_{FS} \;\;.
\label{eq:netbr}
\end{equation}
A typical range for the cold nuclear matter transport coefficient 
for gluons is given by $({\mu^2}/{\lambda}_g)_{eff,\;IS \approx FS} = 
2\times 0.1$~GeV$^2$/fm - $2\times 0.15$~GeV$^2$/fm.
For final state scattering in a 1+1D Bjorken expanding quark-gluon plasma 
the broadening $(({\mu^2}/{\lambda})_{eff} \langle L \rangle_{FS})$ can 
be evaluated from Eq.(\ref{broad}) with $dN^g/dy=1150$, 
consistent with the inclusive hadron suppression pattern in 
Fig~\ref{fig1}.

\begin{figure}
  \includegraphics[width=2.7in,height=2.8in]{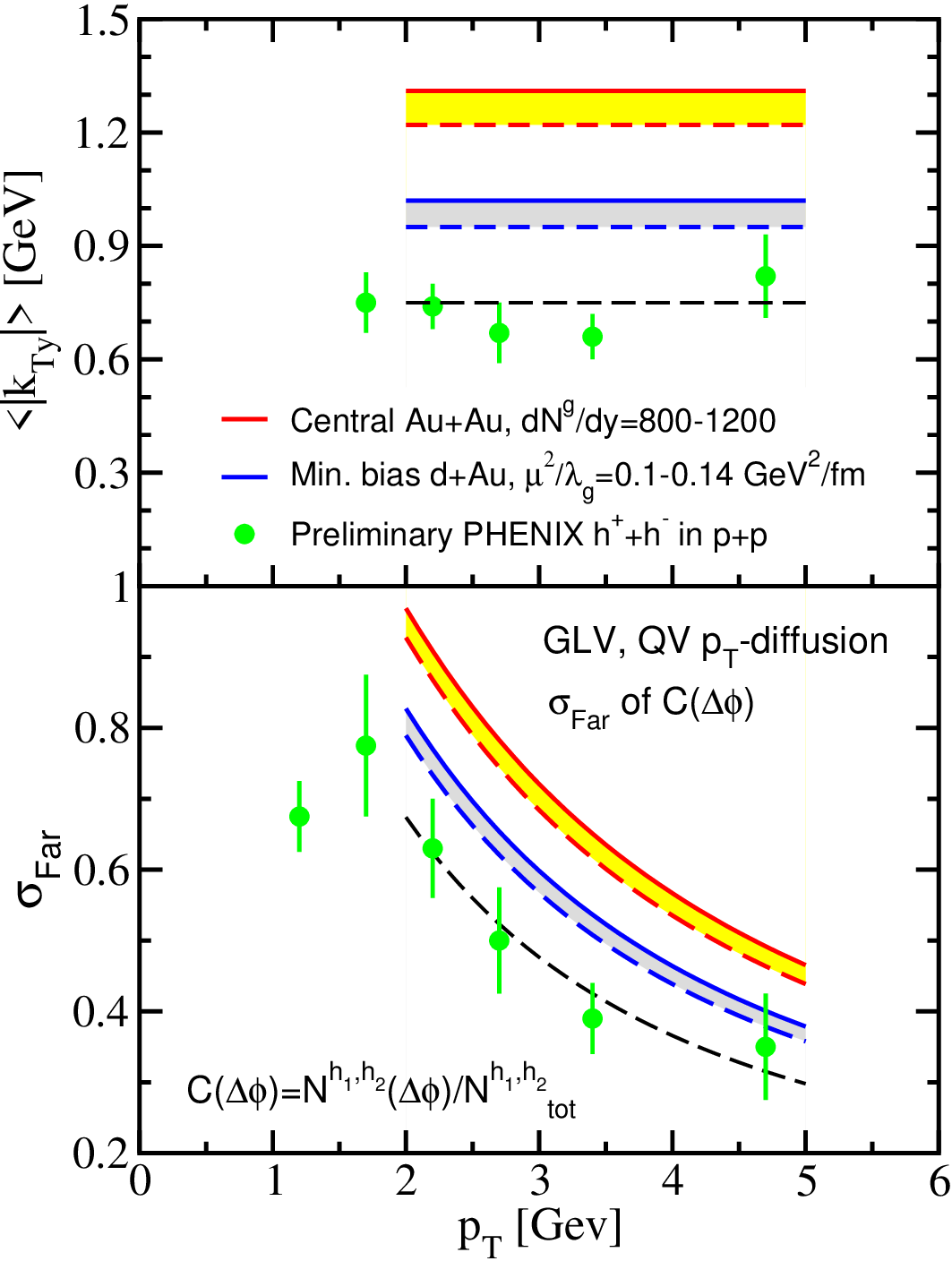}
\hspace*{0.3cm}
  \includegraphics[width=2.7in,height=2.8in]{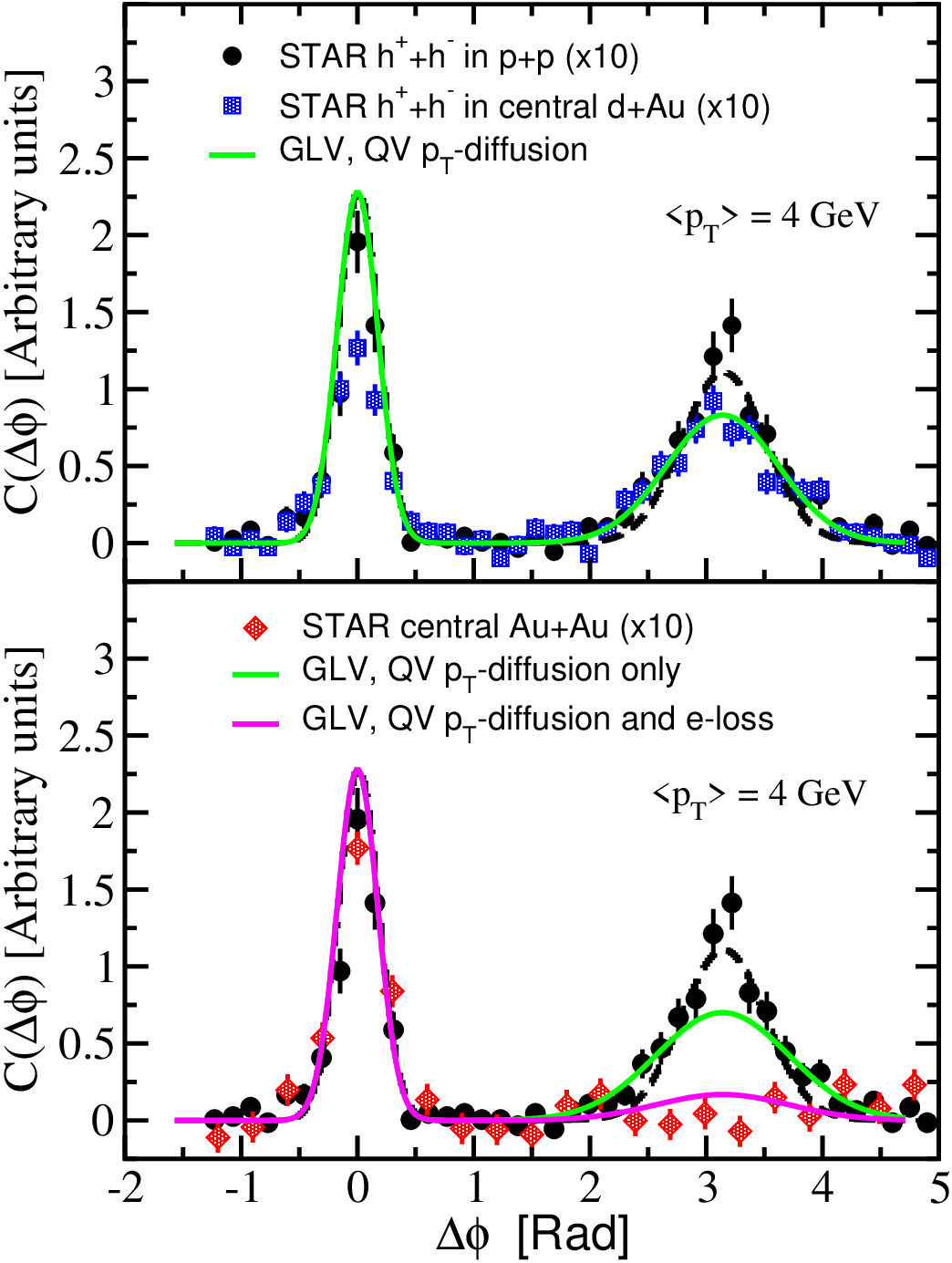}
\vspace*{0.3cm}
  \caption{Left panel: predicted enhancement of 
$\langle |{\bf k}_{T\;y}| \rangle$ and  $\sigma_{Far}$ in minimum bias
$d+Au$ and central $Au+Au$ reactions at RHIC from 
$p_T$-diffusion~\cite{Qiu:2003pm}. Preliminary $p+p$ data is from 
PHENIX~\cite{Adler:2002tq}.  Right panel:  the broadening of the 
far-side di-hadron correlation function in central $d+Au$ and $Au+Au$ 
compared to scaled (x10) STAR data~\cite{Adler:2003ii}. In the 
bottom right panel the broadening with and without suppression, 
approximately given by $R_{AA}$ from Fig.~\ref{fig1}, are shown.}
\label{fig2}
\end{figure}

Figure~\ref{fig2} shows two measures of the predicted increase in 
di-jet acoplanarity for minimum bias $d+Au$ and central $Au+Au$ 
reactions~\cite{Qiu:2003pm}: 
$\langle |{\bf k}_{T\;y}| \rangle = 
\sqrt{\langle  {\bf k}_T^2 \rangle_{1\, parton}/\pi}$, 
($\langle {\bf k}_T^2 \rangle_{1\, parton} = 
\langle {\bf k}_T^2 \rangle/2$) and the away-side width $\sigma_{Far}$  
of the di-hadron correlation function  $C(\Delta \phi) = 
N^{h_1,h_2}(\Delta \phi)/N_{tot}^{h_1,h_2}$. $C(\Delta \phi)$ is 
approximated by near-side and far-side Gaussians for a symmetric 
$p_T^{h1} \approx p_T^{h2}$ case and the vacuum widths are  
taken from PHENIX~\cite{Adler:2002tq}. 
In the right panel of Fig.~\ref{fig2} di-hadron correlations in 
$d+Au$ are shown to be qualitatively similar to the $p+p$ case 
and in agreement with STAR measurements~\cite{Adler:2003ii}.  
In $Au+Au$ reactions at RHIC di-jet acoplanarity is 
noticeably larger, but this effect alone does not lead to the 
reported  disappearance of the back-to-back 
correlations~\cite{Adler:2002tq}. To first approximation the 
coefficient of the away-side Gaussian
(the area under $C(\Delta \phi)$, $\Delta \phi > \pi/2$),  
is determined by jet energy loss and given by $R_{AA} \propto 
N_{part}^{2/3}$. Broadening with and  without away-side quenching 
is shown the bottom right panel of Fig.~\ref{fig2}. 
Combined $d+A$u and $Au+Au$ experimental data in Fig.~\ref{fig2} 
also rule out the existence of monojets at RHIC. For further 
discussion on di-hadron correlations see~\cite{Qiu:2003pm,Hirano:2003hq}.

In summary, evidence from jet tomography~\cite{Vitev:2002pf,Arleo:2002kh}, 
relativistic hydrodynamics~\cite{Huovinen:2003fa}
and parton cascade models~\cite{Zhang:1999bd} is in strong support 
of the creation of a deconfined phase of QCD at RHIC with initial 
energy density $\sim 20$~GeV/fm$^3$, more  than 100 times the  
$1/7$~GeV/fm$^3$ density of cold nuclear matter.

\begin{theacknowledgments}
Useful discussion with J.~Qiu is acknowledged. Thanks to 
I.~Bearden, J.~Dunlop, B.~Jacak, J. Klay, M.~Miller and G.~Roland 
for help with experimental data. This work is supported by  
the U.S. Department of Energy  under Contract No. DE-FG02-87ER40371.
\end{theacknowledgments}

\bibliographystyle{aipproc}   % if natbib is available

\end{document}